\documentclass[prd,12pt,tightenlines,nofootinbib]{revtex4}

\usepackage[english]{babel}
\usepackage{graphicx}
\usepackage{psfrag}
\usepackage{amsmath}
\usepackage{amssymb}
\usepackage{bbm}
\usepackage{slashed}
\usepackage{verbatim}
\usepackage[hypertex]{hyperref}

\newcommand{\be}{\begin{equation}}
\newcommand{\ee}{\end{equation}}
\newcommand{\bea}{\begin{eqnarray}}
\newcommand{\eea}{\end{eqnarray}}
\newcommand{\bean}{\begin{eqnarray*}}
\newcommand{\eean}{\end{eqnarray*}}

\renewcommand{\b}{\langle}
\newcommand{\ket}{\rangle}

\newcommand{\irm}{{\rm i}}
\newcommand{\e}{{\rm e}}

\renewcommand{\d}{{\rm d}}
\newcommand{\cl}[1]{{\mathcal #1}}

\renewcommand{\v}[1]{\vec{#1}}

\newcommand{\ds}{\displaystyle}

\newcommand{\bZ}{\mathbb{Z}}
\newcommand{\bN}{\mathbb{N}}
\newcommand{\bC}{\mathbb{C}}
\newcommand{\bR}{\mathbb{R}}

\newcommand{\bH}{\mathbb{H}}

\newcommand{\clH}{\cl{H}}

\newcommand{\clD}{\cl{D}}

\newcommand{\clJ}{\cl{J}}
\newcommand{\clC}{\cl{C}}

\newcommand{\eq}[1]{(\ref{#1})}
\renewcommand{\sec}[1]{sec.\ \ref{#1}}

\newcommand{\threevector}[3]
{\left(\begin{array}{c} #1 \\ #2 \\ #3 \end{array}\right)}

\newcommand{\qed}{\nobreak \ifvmode \relax \else
      \ifdim\lastskip< 1 em \hskip-\lastskip
      \hskip1.0em plus0em minus0.5em \fi \nobreak
      \vrule height0.75em width0.75em depth0 em\fi}

\newcommand{\xib}{\overline{\xib}}

\newcommand{\Hphys}{\clH_{\mathrm{phys}}}
\newcommand{\Pphys}{P_{\mathrm{phys}}}
\newcommand{\SLC}{\mathrm{SL(2,\bC)}}

\begin{document}

\title{Spin foams with timelike surfaces}

\author{Florian Conrady}
\email{fconrady@perimeterinstitute.ca}
\affiliation{Perimeter Institute for Theoretical Physics, Waterloo, Ontario, Canada}

\begin{abstract}
Spin foams of 4d gravity were recently extended from complexes with purely spacelike surfaces
to complexes that also contain timelike surfaces.
In this article, we express the associated partition function in terms of vertex amplitudes
and integrals over coherent states.
The coherent states are characterized by unit 3--vectors which
represent normals to surfaces and lie either in the 2--sphere or the 2d hyperboloids.
In the case of timelike surfaces, a new type of coherent state is used
and the associated completeness relation is derived.

It is also shown that the quantum simplicity constraints can be deduced
by three different methods: by weak imposition of the constraints, by restriction of
coherent state bases and by the master constraint.
\end{abstract}

\maketitle

\section{Introduction}
\label{introduction}

The theory of spin foams rests on the idea that quantum spacetime is
a process of transitions between quanta of geometry. In a sense, one
can think of a spin foam as a Feynman diagram
\cite{ReisenbergerRovelliFeynman1,ReisenbergerRovelliFeynman2,Rovellibook,Oritigroupfieldtheoryapproach}.
Each vertex of this diagram
corresponds to a quantum 4--simplex and an edge describes the
propagation of a tetrahedron in one 4--simplex to a tetrahedron of
another 4--simplex. The end of an edge carries a quantum label that
denotes a quantum state of a tetrahedron---similar to a Fock state
in quantum field theory. These labels are, in turn, made up of
quantum numbers for triangles that form the tetrahedron. Typically,
the quantum numbers are spins, hence the name spin foam \cite{Baezspinfoammodels}.
This
covariant picture is complemented by the canonical framework of loop
quantum gravity, which provides an operator formalism for quantum
tetrahedra and more general 3--geometries \cite{Rovellibook,Thiemannbook}.

In recent years, considerable progress was made in defining spin
foams and their dynamics. The so--called simplicity constraints play
a key role in these developments. They have the purpose of
constraining states of a topological theory (BF theory) to states of
4d quantum gravity (where the B field is replaced by tetrads). In
the same way that QCD has different lattice actions, the resulting
quantum gravity has different variants: originally the BC model
\cite{BarrettCrane} and more recently the EPRL \cite{EPRL} and FK model \cite{FK,Livine_New,Livine_Consistently}.

The EPRL model comes in two versions, one for Riemannian geometries and one for Lorentzian signature.
The Lorentzian model is subject to the restriction that tetrahedra and triangles are spacelike.
As a result, these spin foams correspond to a relatively special class of
Lorentzian triangulations. In particular, boundary hypersurfaces are always
spacelike. In a recent paper by the author and J.\ Hnybida this theory
was extended to include also tetrahedra with Lorentzian signature and hence
timelike triangles \cite{CHtimelike}. Like for spacelike surfaces, the spectrum of timelike
areas turns out to be discrete.

This extension is certainly natural from a covariant perspective.
If one can implement Lorentzian tetrahedra, there is a priori
no reason to forbid them. The inclusion of such tetrahedra has the advantage of permitting
timelike boundaries, which is not possible otherwise.
By using general triangualations one might also avoid artifacts or distortions
that could arise when the triangulations are restricted\footnote{Compare this with the issue of ``fatness'' and convergence in Riemannian Regge calculus \cite{CheegerMullerSchrader}.}.

Is the restriction to Euclidean tetrahedra required or favored from a Hamiltonian point of view?
In the examples we know of, the transition from space to spacetime
leads to 4d lattices that include timelike or null edges. This is the case
in causal dynamical triangulations \cite{AmbjornetalDynamicallytriangulating}, for instance, and in
evolution algorithms for classical Lorentzian Regge calculus \cite{evolutionscheme}.
It may be possible to evolve on triangulations with purely spacelike edges,
but we are not aware of any examples for this.
The Hamiltonian approach to Lorentzian spin foams \cite{Hanpathintegral,HanThiemannrelation}
results in sequences of 3d spatial lattices,
so this cannot (yet) be directly compared with the 4d triangulations discussed here.

In the present article, we address some of the points that were left open in ref.\ \cite{CHtimelike}.
Firstly, we express the spin foam theory in terms of vertex amplitudes that have coherent
states as boundary data.
For this we derive completeness relations for a new type of coherent state
needed to describe timelike triangles.
Secondly, we show that the quantum simplicity constraints can be obtained in three different ways:
1.\ by weak imposition of the constraints, 2.\ by restriction of a coherent state basis (a more formal
version of the argument given in \cite{CHtimelike}), and 3.\ by the master constraint.
The consistency of these methods provides an additional check on the correctness of the constraints.

The paper is organized as follows.
In section \ref{reduction} we briefly recall some facts about
representation theory of $\mathrm{SL(2,\bC)}$, SU(2) and SU(1,1)
that we need in the remainder of the text. In \sec{completenessrelations} we state the known completeness
relations for SU(2) and the discrete series of SU(1,1), and we derive the one for the new coherent state
introduced in \cite{CHtimelike}. The three derivations of the quantum simplicity constraints
are described in \sec{threeways}. Finally, the completeness relations are used to write the spin foam model
as a multiple integral over vertex amplitudes (\sec{coherentstateintegral}).

\section{SU(2) and SU(1,1) reduction of $\mathrm{SL(2,\bC)}$ representations}
\label{reduction}

\setlength{\jot}{0.5cm}

This section summarizes a number of facts about irreducible representations of $\mathrm{SL(2,\bC)}$, SU(2) and SU(1,1)
that are essential for the definition of the spin foam model. $\mathrm{SL(2,\bC)}$ has the generators
\be
J^i = \sigma^i/2\,,\qquad K^i = \irm\sigma^i/2\,,\qquad i = 1,2,3\,,
\ee
and the subgroups SU(2) and SU(1,1) are generated by $J^1, J^2, J^3$ and $J^3, K^1, K^2$ respectively.
Unitary irreps of $\mathrm{SL(2,\bC)}$ are labelled by pairs $(\rho,n)$, where $\rho\in\bR$ and $n\in\bZ_+$.
The associated Hilbert space and representation matrices are denoted by $\clH_{(\rho,n)}$ and $D^{(\rho,n)}(g)$,
$g\in \mathrm{SL(2,\bC)}$, respectively.
There are two Casimirs, given by
\bea
C_1 &=& 2\left(\v{J}^2 - \v{K}^2\right) = \frac{1}{2} (n^2 - \rho^2 - 4)\,, \\
C_2 &=& -4 \v{J}\cdot \v{K} = n\rho\,.
\eea
Both the SU(2) and SU(1,1) irreps are built from eigenstates $|j\, m\ket$ of $J^3$:
\be
\b j\, m | j\, m'\ket = \delta_{mm'}\,,\qquad J^3\, |j\, m\ket = m |j\, m\ket\,.
\ee
In the case of SU(2), the irreps are labelled by the Casimir $\v{J}^2$:
\be
\v{J}^2\, |j\, m\ket = j(j+1) |j\, m\ket\,,\qquad\mbox{where $j$ = $0$, $\frac{1}{2}$, $1$, $\frac{3}{2}$, \ldots}
\ee
$\clD_j$ stands for the Hilbert space of spin $j$.
Unitary irreps of SU(1,1) have the Casimir $Q = (J^3)^2 - (K^1)^2 - (K^2)^2$ and split into two classes,
the discrete series and the continuous series. For the discrete series, one has
\be
Q\, |j\, m\ket = j(j-1) |j\, m\ket\,,\qquad\mbox{where $j$ = $\frac{1}{2}$, $1$, $\frac{3}{2}$, \ldots}
\ee
and the eigenvalues of $J^3$ assume the values
\be
m = j,\; j+1,\; j+2,\; \ldots\qquad\mbox{or}\qquad m = -j,\; -j-1,\; -j-2,\; \ldots
\ee
We denote the irrep consisting of states $|j\, m\ket$ with $m \gtrless 0$ by $\clD^\pm_j$.

In the case of the continuous series, the Casimir takes on continuous values:
\be
Q\, |j\, m\ket = j(j+1) |j\, m\ket\,, \qquad\mbox{where $j = -\frac{1}{2} + \irm s$,\quad $0 < s < \infty$,}
\ee
and
\be
m = 0,\, \pm 1,\, \pm 2,\, \ldots\qquad\mbox{or}\qquad m = \pm\frac{1}{2},\, \pm\frac{3}{2},\, \ldots
\ee
Irreps of this series are denoted by $\clC^\epsilon_s$. The label $\epsilon = 0, \frac{1}{2}$ designates the irreps with integer $m$
and half--integer $m$ respectively.

\renewcommand{\arraystretch}{2}
Clearly, every unitary irrep of $\mathrm{SL(2,\bC)}$ defines a representation of its subgroups SU(2) and SU(1,1).
However, these representations are reducible. As a result, the Hilbert space $\clH_{(\rho,n)}$ splits into
a direct sum of irreps of SU(2), or a direct sum of irreps of SU(1,1).
The SU(2) decomposition is given by the following isomorphism and completeness relation:
\be
\clH_{(\rho,n)} \simeq \bigoplus\limits_{j = n/2}^{\infty} \clD_j\,,\qquad
\mathbbm{1}_{(\rho,n)} = \sum\limits_{j = n/2}^\infty \sum_{m=-j}^j \left|\Psi_{j\, m}\right\ket \left\b\Psi_{j\, m}\right|\,.
\label{decompositionSU(2)}
\ee
The states $|\Psi_{j\, m}\ket$ form the so--called canonical basis of $\clH_{(\rho,n)}$.
For fixed spin $j$ and $m = -j,\ldots, j$, they span a subspace of $\clH_{(\rho,n)}$ that is isomorphic
to $\clD_j$. The SU(1,1) reduction can be formally written as
\be
\clH_{(\rho,n)} \quad\simeq\quad
\left(\bigoplus\limits_{j > 1/2}^{n/2} \clD^+_j \oplus\!\!\!\!\! \int\limits_0^{\;\;\;\;\;\infty\; \oplus} \!\!\!\!\!\d s\; \clC^\epsilon_s\right)
\oplus
\left(\bigoplus\limits_{j > 1/2}^{n/2} \clD^-_j \oplus\!\!\!\!\! \int\limits_0^{\;\;\;\;\;\infty\; \oplus} \!\!\!\!\!\d s\; \clC^\epsilon_s\right)\,.
\label{decompositionSU(1,1)}
\ee
The precise meaning of this statement is encoded in the completeness relation
\be
\mathbbm{1}_{(\rho,n)} =
\sum\limits_{j > 1/2}^{n/2} \sum_{m=j}^\infty \left|\Psi^+_{j\, m}\right\ket \left\b\Psi^+_{j\, m}\right|
+ \sum\limits_{j > 1/2}^{n/2} \sum_{-m=j}^\infty \left|\Psi^-_{j\, m}\right\ket \left\b\Psi^-_{j\, m}\right|
+ \sum_{\alpha = 1,2}
\int\limits_0^\infty \d s\;\mu_\epsilon(s) \sum\limits_{\pm m = \epsilon}^\infty \left|\Psi^{(\alpha)}_{s\, m}\right\ket \left\b\Psi^{(\alpha)}_{s\, m}\right|\,.
\label{completenessrelationSU(1,1)}
\ee
Here, $|\Psi^\pm_{j\, m}\ket$ and $|\Psi^{(\alpha)}_{s\, m}\ket$ are states that correspond to states of the discrete and continuous series respectively.
The sum over $j$ extends over values such that $j - n/2$ is integral. Moreover, $\epsilon$ has a value such that $\epsilon - n/2$ is an integer.
The measure factors are
\be
\mu_\epsilon(s) = \left\{
\begin{array}{ll}
\ds 2 s\tanh(\pi s)\,, & \epsilon = 0\,, \\
\ds 2 s\coth(\pi s)\,, & \epsilon = 1/2\,.
\end{array}
\right.
\ee
Note that the irreps $\clD^\pm_{1/2}$ do not appear in the decomposition \eq{decompositionSU(1,1)}.
As a result, the discrete series is absent for $n = 0$ and $n = 1$.

When $\mathrm{SL(2,\bC)}$ is restricted to SU(1,1), the states $\left|\Psi^\pm_{j\, m}\right\ket$ furnish irreducible representations
that are isomorphic to those of the discrete series:
\bea
&& \big\b\Psi^{\pm}_{j\, m'} \big|\Psi^\pm_{j\, m}\big\ket = \delta_{m'm}\,, \\
&& \big\b\Psi^{\pm}_{j\, m'} \big| D^{(\rho,n)}(g) \big|\Psi^\pm_{j\, m}\big\ket = \b j\, m'| D^j(g) |j\, m\ket\quad\mbox{for $g\in\mathrm{SU(1,1)}$.}
\label{reducedrepresentation}
\eea

With regard to the continuous series, the situation is more subtle. Firstly, the continuous series states $\left|\Psi^{(\alpha)}_{s\, m}\right\ket$ appear twice,
which is indicated by the index $\alpha = 1,2$. Moreover, these states are not normalizable:
\be
\big\b\Psi^{(\alpha')}_{s'\, m'}\big|\Psi^{(\alpha)}_{s\, m}\big\ket = \frac{\delta(s'-s)}{\mu_\epsilon(s)}\,\delta_{\alpha'\alpha}\,\delta_{m'm}\,
\label{orthogonalitycontinuousseries}
\ee
Thus, the analog of eq.\ \eq{reducedrepresentation} requires an integration over $s$:
\be
\int\limits_0^\infty \d s'\;\mu_\epsilon(s') \big\b\Psi^{(\alpha)}_{s'\, m'}\big| D^{(\rho,n)}(g)\big|\Psi^{(\alpha)}_{s\, m}\big\ket
= \b j\, m'| D^j(g) |j\, m\ket\quad\mbox{for $g\in\mathrm{SU(1,1)}$.}
\label{isomorphismcontinuousseries}
\ee
The expansions \eq{decompositionSU(2)} and \eq{completenessrelationSU(1,1)} follow from the
the Plancherel decomposition of functions on SU(2) and SU(1,1) respectively (see chapter 3 and 7 in \cite{Ruhl} and sec.\ 12 and 13 in \cite{Bargmann}).
In the case of SU(1,1),
the orthogonality relations of matrix elements take the form
\be
\int\limits_{\mathrm{SU(1,1)}}\!\!\!\! \d g\;
D^{j' *}_{m'_1 m'_2}(g) D^j_{m_1 m_2}(g)
=
\frac{1}{2j - 1} \delta_{j' j}\, \delta_{m'_1 m_1} \delta_{m'_2 m_2}
\label{orthogonalitymatrixdiscrete}
\ee
for the discrete series\footnote{When comparing with Ruhl's book, it is important to know that there is a difference in convention:
in the case of the discrete series, our representation matrix $D^j$ is the same as Ruhl's matrix $D^{j-1}$ (see p.\ 227, \cite{Ruhl}).}
with $j > 1/2$, and for the continuous series
\bea
&& \int\limits_{\mathrm{SU(1,1)}}\!\!\!\! \d g\; \int\limits_0^\infty \d s'\; \psi'^*(s') \int\limits_0^\infty \d s\; \psi(s)\;
D^{j' *}_{m'_1 m'_2}(g) D^j_{m_1 m_2}(g) \nonumber \\
&&=
\left(\int\limits_0^\infty \d s\; \frac{1}{\mu_{\epsilon}(s)}\psi'^*(s)\psi(s)\right) \delta_{m'_1 m_1} \delta_{m'_2 m_2}\,,
\label{orthogonalitymatrixcontinuous}
\eea
where $\psi$ and $\psi'$ are square--integrable functions of $s$.

\section{Completeness relations of SU(1,1) coherent states}
\label{completenessrelations}

The quantization of the simplicity constraint leads altogether to three
types of constraints that reflect three different possibilities at the
classical level: 1.\ a spacelike triangle in a tetrahedron with a timelike normal $U$,
2.\ a spacelike triangle in a tetrahedron with a spacelike normal $U$, and
3.\ a timelike triangle in a tetrahedron with a  spacelike normal $U$.
The spin foam model in \cite{CHtimelike} covers all three cases and represents,
in this sense, a quantization of general Lorentzian geometries.
Its partition function is defined by means of projectors that project onto
those irreps permitted by the three kinds of constraints.

Ref.\ \cite{CHtimelike} provided also a definition of the coherent state vertex amplitude.
The coherent states encode the boundary geometry of a 4--simplex dual to the vertex.
They are labelled by 3--vectors $\v{N}$ that represent the unit normals of triangles
within the 3d boundary of the 4--simplex.
In \cite{CHtimelike} the vertex amplitude was not yet used for defining the partition function. In order
to do so one has to express the aforementioned projectors
in terms of coherent states. That is, one needs completeness relations
that resolve the identity on a given irrep as an integral over coherent states.

For SU(2) and the discrete series of SU(1,1) such completeness relations
are already known \cite{Perelomov}. However, in the case of the continuous series, we employ
a new class of coherent states whose completeness has not been proven so far.
In this section, we will deliver this proof by using orthogonality relations of SU(1,1) matrix elements.
Since the states are not normalizable, a smearing procedure is necessary.
In section \ref{coherentstateintegral}, this result will be applied to express the
spin foam sum of \cite{CHtimelike} as a multiple integral over vertex amplitudes.

\subsection{SU(2) and discrete series of SU(1,1)}

Before coming to the continuous series, we recall the completeness relations for SU(2) and
the discrete series of SU(1,1) \cite{Perelomov} and state them in a form suitable for this
paper. For SU(2) one has the completeness relation
\be
\mathbbm{1}_j = (2j + 1) \int\limits_{\mathrm{SU(2)}}\!\!\! \d g\; |j\, g\ket \b j\, g| = (2j + 1) \int\limits_{S^2} \d^2 N\; |j\v{N}\ket
\ee
where $|j\, g\ket$ and $|j\v{N}\ket$ are the well--known coherent states of SU(2).
The SU(2) measure is the normalized Haar measure and $\d^2 N$ denotes the
normalized measure of the 2--sphere. When translated to the canonical basis of $\clH_{(\rho,n)}$, these formulae read
\be
P_j = (2j + 1) \int\limits_{\mathrm{SU(2)}}\!\!\! \d g\; \left|\Psi_{j\, g}\right\ket \left\b\Psi_{j\, g}\right|
= (2j + 1) \int\limits_{S^2} \d^2 N\; \left|\Psi_{j\v{N}}\right\ket \left\b\Psi_{j\v{N}}\right|\,.
\label{completenessSU(2)}
\ee
$P_j$ is the projector from $\clH_{(\rho,n)}$ to the subspace isomorphic to $\clD_j$.

In the case of the discrete series, SU(1,1) is suitably parametrized by
\be
g = \e^{-\irm \varphi J^3}\, \e^{-\irm u K^1}\, \e^{\irm \psi J^3}\,,\qquad -\pi < \varphi \le \pi\,,\quad 0 \le u < \infty\,, \quad -2\pi < \psi \le 2\pi\,.
\label{parametrizationdiscrete}
\ee
For the SU(1,1) measure, we adopt the same normalization as in ref.\ \cite{Bargmann}. In terms of the coordinates \eq{parametrizationdiscrete}, this gives
\be
\d g = \frac{1}{(4 \pi)^2}\, \sinh u\, \d\varphi\,\d u\, \d\psi\,.
\label{measureSU(1,1)}
\ee
The components $\bH_\pm = \{\v{N}\,|\, \v{N}^2 = 1\,, N^0 \gtrless 0\}$ of the timelike two--sheeted hyperboloid $\bH_+ \cup \bH_-$ are isomorphic to
the quotient $\mathrm{SU(1,1)}/\mathrm{U(1)}$ and can be parametrized by
\be
\v{N} = \pm (\cosh u,\sin\varphi \sinh u,\cos\varphi \sinh u)\,,\qquad -\pi < \varphi \le \pi\,,\quad 0 \le u < \infty\,,
\ee
with the measure
\be
\d^2 N = \frac{1}{4\pi}\, \sinh u\, \d\varphi\, \d u\,.
\ee
For the irreps $D^\pm_j$, the coherent states are defined by
\bea
|j\, g\ket_\pm &\equiv& D^j(g) |j \pm\!j\ket\,,\qquad g\in \mathrm{SU(1,1)}\,, \\
|j\,\v{N}\ket &\equiv& D^j(g(\v{N})) |j \pm\!j\ket\,,\qquad \v{N}\in \bH_\pm\,,
\eea
where $g(\v{N})$ is the SU(1,1) element determined by
\be
g = \e^{-\irm \varphi J^3}\, \e^{-\irm u K^1}\,.
\ee
Then, the completeness relation for $\clD^\pm_j$, $j > 1/2$, can be written as
\be
\mathbbm{1}^\pm_j = (2j - 1) \int\limits_{\mathrm{SU(1,1)}}\!\!\!\! \d g\; |j\,g\ket_\pm \b j\, g|_\pm
= (2j - 1) \int\limits_{\bH_\pm} \d^2 N\; |j\,\v{N}\ket \b j\,\v{N}|\,.
\ee
It can be derived from the orthogonality relation \eq{orthogonalitymatrixdiscrete} between matrix elements of SU(1,1) \cite{Bargmann}.
At the level of the $\mathrm{SL(2,\bC)}$ irrep $\clH_{(\rho,n)}$, the coherent states correspond to
\bea
\left|\Psi^\pm_{j\, g}\right\ket &\equiv& D^{(\rho,n)}(g) \left|\Psi^\pm_{j\,\pm j}\right\ket\,,\qquad g\in \mathrm{SU(1,1)}\,, \\
\left|\Psi_{j\v{N}}\right\ket &\equiv& D^{(\rho,n)}(g(\v{N})) \left|\Psi^\pm_{j\,\pm j}\right\ket\,,\qquad \v{N}\in\bH_\pm\,,
\eea
and the completeness relation becomes
\be
P^\pm_j
= (2j - 1) \int\limits_{\mathrm{SU(1,1)}}\!\!\!\! \d g\; \left|\Psi^\pm_{j\, g}\right\ket \left\b\Psi^\pm_{j\, g}\right|
= (2j - 1) \int\limits_{\bH_\pm} \d^2 N\; \left|\Psi_{j\v{N}}\right\ket \left\b\Psi_{j\v{N}}\right|\,.
\label{completenessdiscrete}
\ee
Here, $P^\pm_j$ is the projector from $\clH_{(\rho,n)}$ onto the subspace isomorphic to $\clD^\pm_j$.

\subsection{Continuous series}

In the case of the continuous series, we build coherent states from eigenstates of $K^1$ \cite{CHtimelike}.
For this reason, it is practical to parametrize SU(1,1) by
\be
g = \e^{-\irm \varphi J^3}\, \e^{-\irm t K^2}\, \e^{\irm u K^1}\,,\qquad -2\pi < \varphi \le 2\pi\,,\quad -\infty < t, u < \infty\,,
\label{parametrizationcontinuous}
\ee
where the right--most factor is generated by $K^1$ \cite{Lindblad}.
In these coordinates, the measure \eq{measureSU(1,1)} reads
\be
\d g = \frac{1}{(4 \pi)^2}\, \cosh t\, \d\varphi\, \d t\,\d u\,.
\ee
The relevant hyperboloid is now the spacelike single--sheeted hyperboloid $\bH_{\mathrm{sp}} = \{\v{N}\,|\, \v{N}^2 = -1\}$.
It is isomorphic to the quotient $\mathrm{SU(1,1)} / (\mathrm{G_1}\otimes\bZ_2)$, where $G_1$ is the one--parameter subgroup
generated by $K^1$. We coordinatize $\bH_{\mathrm{sp}}$ by
\be
\v{N} = (-\sinh t,\cos\varphi \cosh t,\sin\varphi \cosh t)\,,\qquad -\pi < \varphi \le \pi\,,\quad -\infty < t < \infty\,,
\ee
and fix the measure to be
\be
\d^2 N = \frac{1}{4\pi}\, \cosh t\, \d\varphi\, \d t\,.
\label{measureNcontinuous}
\ee
In the irrep $C^\epsilon_s$, eigenstates of $K^1$ with eigenvalue $\lambda$ are denoted by $|j\,\lambda\,\sigma\ket$.
The spectrum of $K^1$ is the real line and it is two--fold degenerate, so there is an additional label $\sigma = \pm$ (see \cite{LindbladNagel} for more details).
Like momentum eigenstates, these states are not normalizable:
\be
\b j\,\lambda'\,\sigma' |j\,\lambda\,\sigma\ket = \delta(\lambda' - \lambda) \delta_{\sigma'\sigma}
\label{orthogonalityjlambda}
\ee
Our coherent states result from SU(1,1) transformations of the reference state $|j\,s\,+\ket$ with eigenvalue $\lambda = s$:
\bea
|j\,g\ket_{\mathrm{sp}} &\equiv& D^j(g) |j\, s\, +\ket\,,\qquad g\in \mathrm{SU(1,1)}\,, \\
|j\,\v{N}\ket &\equiv& D^j(g(\v{N})) |j\, s\, +\ket\,.
\eea
$g(\v{N})$ is the SU(1,1) element determined by
\be
g = \e^{-\irm \varphi J^3}\, \e^{-\irm t K^2}\,.
\ee
Like for SU(2) and the discrete series of SU(1,1), the completeness relation can be derived from
the orthogonality of matrix elements (see eq.\ \eq{orthogonalitymatrixdiscrete} and \eq{orthogonalitymatrixcontinuous}).
In the case of the continuous series, these matrix elements are not normalizable and a smearing in $s$ is required.
For this reason, we do not resolve the identity on a single irrep $C^\epsilon_s$. Instead we consider a projector
from $\clH_{(\rho,n)}$ to states $|\Psi^{(\alpha)}_{s\, m}\ket$ which permits a range of values of $s$, defined
by a suitable wavefunction. Furthermore, eigenstates of $K^1$ are not normalizable, so we also need a smearing over
eigenvalues $\lambda$.

For the smearing, we choose the function
\be
f_\delta(x) =
\left\{
\begin{array}{ll}
1\,, & |x| \le \delta/2\,, \\
0\,, & |x| > \delta/2\,.
\end{array}
\right.
\ee
The smeared projector is specified by
\be
P^\epsilon_s(\delta) = \sum_{\alpha = 1,2} \sum_{\pm m = \epsilon} \int\limits_0^\infty \d s'\; \mu_\epsilon(s')\, f_\delta(s'-s)
\left|\Psi^{(\alpha)}_{s'\, m}\right\ket \left\b\Psi^{(\alpha)}_{s'\, m}\right|
\label{projector_s}
\ee
It projects $\clH_{(\rho,n)}$ onto the subspace of wavefunctions with support on the interval $[s-\delta/2,s+\delta/2]$. Note that this
is slightly different from the projector chosen in \cite{CHtimelike}.

The smeared coherent states are defined by
\be
\left|\Psi^{(\alpha)}_{s\, g\, \delta}\right\ket \equiv
\int\limits_0^\infty \d s'\; \mu_\epsilon(s')\, f_\delta(s'-s)
\int\limits_{-\infty}^\infty \d\lambda\; \frac{1}{\sqrt{\delta}}
f_\delta(\lambda - s)\,
D^{(\rho,n)}(g) \left|\Psi^{(\alpha)}_{s'\,\lambda\,+}\right\ket\,,\quad g\in \mathrm{SU(1,1)}\,.
\ee
With these states, the projector \eq{projector_s} can be expressed as
\be
P^\epsilon_s(\delta) =
\sum_{\alpha = 1,2}\; \int\limits_{\mathrm{SU(1,1)}}\!\!\!\! \d g\; \left|\Psi^{(\alpha)}_{s\, g\, \delta}\right\ket \left\b\Psi^{(\alpha)}_{s\, g\, \delta}\right|\,.
\label{completenesscontinuousg}
\ee
The same may be also written as an integral over $\v{N}$. However, due to the smearing, the corresponding states carry additional indices. If we define the smeared coherent states by
\be
\left|\Psi^{(\alpha)}_{j\v{N}\lambda\delta}\right\ket \equiv
\int\limits_0^\infty \d s'\; \mu_\epsilon(s')\, f_\delta(s'-s)\,
D^{(\rho,n)}(g(\v{N})) \left|\Psi^{(\alpha)}_{s'\,\lambda\,+}\right\ket\,,\quad \v{N}\in\bH_{\mathrm{sp}}\,,
\ee
the completeness relation becomes
\be
P^\epsilon_s(\delta) =
\sum_{\alpha = 1,2}\; \int\limits_{\bH_{\mathrm{sp}}} \d^2 N\;
\int\limits_{-\infty}^\infty \d\lambda\; \frac{1}{\delta} f_\delta(\lambda - s)\,
\left|\Psi^{(\alpha)}_{j\v{N}\lambda\delta}\right\ket \left\b\Psi^{(\alpha)}_{j\v{N}\lambda\delta}\right|\,.
\label{completenesscontinuousN}
\ee
Both eq.\ \eq{completenesscontinuousg} and \eq{completenesscontinuousN} are proven in appendix \ref{proof}.
Together eqns.\ \eq{completenessdiscrete} and \eq{completenesscontinuousg} yield the following completeness relation for
the entire $\SLC$ representation space $\clH_{(\rho,n)}$:
\bea
\mathbbm{1}_{(\rho,n)} &=&
\sum\limits_{j > 1/2}^{n/2}\,
(2j - 1)\!\!\!\! \int\limits_{\mathrm{SU(1,1)}}\!\!\!\! \d g\; \left(\left|\Psi^+_{j\, g}\right\ket \left\b\Psi^+_{j\, g}\right| + \left|\Psi^-_{j\, g}\right\ket \left\b\Psi^-_{j\, g}\right|\right) \nonumber \\
&&
{}+
\frac{1}{\delta}\int\limits_0^\infty \d s\!\!\! \int\limits_{\mathrm{SU(1,1)}}\!\!\!\! \d g
\sum_{\alpha = 1,2}\;\left|\Psi^{(\alpha)}_{s\, g\, \delta}\right\ket \left\b\Psi^{(\alpha)}_{s\, g\, \delta}\right|
\label{totalcompletenessrelationSU(1,1)}
\eea

\section{Three ways to simplicity}
\label{threeways}

The quantum simplicity constraint in ref.\ \cite{CHtimelike} were derived from the
requirement that there exist semiclassical states with simple expectation values and
small uncertainties. In this section, we formalize this derivation and describe it as
a projection from the kinematic to the physical Hilbert space.
We demonstrate furthermore that the same simplicity constraints can be obtained from
two other techniques; from the weak imposition of the constraints via matrix elements
and from the master constraint.
The three types of derivations will be first exemplified in a simple system
(which was already mentioned in \cite{EPRflipped}), and then we will move on to the simplicity
constraints themselves.

In contrast to the gauge and diffeomorphism constraint of gravity, the simplicity constraints
are second--class and they require a different treatment for quantization.
The particular features of second--class constraints are best explained by the following
basic example. Consider a system of two particles whose positions and momenta are constrained
to be identical. The phase space variables are given by coordinates $(q_i,p_i)$, $i = 1,2$,
with Poisson bracket $\{q_i,p_i\} = \delta_{ij}$, and the constraints are
\be
q_1 - q_2 = 0\,,\qquad p_1 - p_2 = 0\,.
\ee
One can make a change of coordinates and use equivalently
\be
q_\pm = \frac{1}{2}\left(q_1 \pm q_2\right)\,,\qquad p_\pm = \frac{1}{2}\left(p_1 \pm p_2\right)
\ee
or complex variables
\be
a_\pm = \frac{1}{\sqrt{2}}\left(p_\pm - \irm q_\pm\right)\,,
\ee
in which case the constraints take the form
\be
q_- = p_- = 0\,,\qquad\mbox{or}\qquad a_- = 0\,.
\ee
Upon quantization, the kinematic Hilbert space $\clH$ is given by the Fock space spanned by states
\be
|n_+\ket\otimes |n_-\ket = (a^\dagger_+)^{n_+} |0\ket \otimes (a^\dagger_-)^{n_-} |0\ket
\ee
where $n_+, n_-\in\bN_0$. Clearly, the physical Hilbert space $\Hphys$ should be isomorphic to the Fock space
of a single degree of freedom, and there are different ways to arrive at this conclusion mathematically.

One strategy is to impose the constraints weakly \cite{EPRflipped,Tatealgebraic}, which is related to the Gupta--Bleuler method
in QED and string theory. If one imposed $q_-|\psi\ket = p_-|\psi\ket = 0$ strongly, there would be no non--trivial solution,
so one requires instead only $a_-|\psi\ket = 0$. It then follows that the full constraint holds weakly in the sense that
\be
\b\varphi| a_- |\psi\ket = \b\varphi| a^\dagger_- |\psi\ket = 0\qquad\mbox{$\forall$ $|\varphi\ket, |\psi\ket\in \Hphys$}\,,
\ee
and $\Hphys$ is spanned by the states $|n_+\ket\otimes |0\ket$, $n_+\in\bN_0$.

Another possibility is to start from an overcomplete basis of coherent states for the full Hilbert space $\clH$ and to restrict this basis,
so that only states in $\Hphys$ remain. In the present case, the kinematic Hilbert space is spanned by coherent states
$|\alpha_+\ket \otimes |\alpha_-\ket$, where $a_\pm|\alpha_\pm\ket = \alpha_\pm |\alpha_\pm\ket$, with the completeness relation
\be
\mathbbm{1}_{\clH} = \frac{1}{\pi^2}\int \d\alpha_+ \int \d\alpha_-\; |\alpha_+\ket \b\alpha_+| \otimes |\alpha_-\ket \b\alpha_-|\,.
\ee
The projector on $\Hphys$ is obtained by restricting the integral to coherent states whose expectation values satisfy the constraint.
That is, $(\b \alpha_+| \otimes \b \alpha_-|)\, a_-\, (| \alpha_+\ket \otimes |\alpha_-\ket) = \alpha_- = 0$ and hence
\be
\Pphys \equiv \frac{1}{\pi}\int \d\alpha_+\; |\alpha_+\ket \b\alpha_+| \otimes |0\ket \b 0|\,.
\ee
The normalization is adjusted to ensure that the constrained integral is a projector.

A third method is based on the so--called master constraint\footnote{For the general idea, see e.g.\ sec.\ 2 in \cite{DittrichThiemannTestingI}.}.
The master constraint is the sum of the squares of the constraints,
here $M = p^2_- + q^2_- = 0$. Up to factor ordering, this constraint becomes
\be
M = a^\dagger_- a_- = \frac{1}{2}\left(p^2_- + q^2_-\right) + \frac{1}{2}
\ee
in the quantum theory. The physical Hilbert space is defined as the subspace of states with minimal eigenvalue w.r.t.\ $M$,
which consists, as before, of the states $|n_+\ket \otimes |0\ket$, $n_+\in\bN_0$.

Below we will translate these three methods to the simplicity constraints of spin foams\footnote{For a more detailed explanation of simplicity constraints, \
see e.g.\ \cite{FK} or sec.\ II in \cite{CHtimelike}.}.
Classically, the variables are given by SO(1,3) bivectors $B$ that are constrained
to be simple. On a simplicial complex, these bivectors are associated to triangles and represent the
area bivectors of the triangles.
Simplicity is encoded by assigning a normal 4--vector $U$ to each tetrahedron
and by requiring that
\be
U\cdot\star B = 0
\label{simplicityconstraint}
\ee
for all four area bivectors of the tetrahedron.
At the classical level, the bivectors are related to bivectors $J = (J^{IJ})$ by
\be
B = \frac{\gamma^2}{\gamma^2 + 1}\left(J - \frac{1}{\gamma} \star J\right)\,.
\label{relationbivectorgenerator}
\ee
The latter correspond to the generators of the Lorentz group in the quantum theory.
$\gamma$ is the Immirzi parameter and we assume that $\gamma > 0$.

\renewcommand{\arraystretch}{1.5}
The tetrahedral normal $U$ is assumed to be timelike or spacelike, and after gauge--fixing
these two possibilities are represented by the values $U = (1,0,0,0)$ and $U = (0,0,0,1)$.
Using the relation \eq{relationbivectorgenerator} the simplicity constraint \eq{simplicityconstraint}
is then expressed in terms of $\SLC$ generators, namely
\be
\v{C} = \v{J} + \frac{1}{\gamma}\v{K} = 0\qquad\mbox{and}\qquad \v{C} = \v{F} + \frac{1}{\gamma}\v{G} = 0
\label{simplicityconstraintsaftergaugefixing}
\ee
for $U = (1,0,0,0)$ and $U = (0,0,0,1)$ respectively, where
\be
\v{F} = \threevector{J^3}{K^1}{K^2}\qquad\mbox{and}\qquad\v{G} = \threevector{K^3}{-J^1}{-J^2}\,.
\ee
This is the form of the constraints that are quantized.
In addition, we also use the diagonal constraint $B\cdot \star B = B^{IJ}(\star B)_{IJ} = 0$, since it is implied by
the simplicity constraint \eq{simplicityconstraint}. The former is first--class and easily implemented,
 as it can be expressed in terms of $\mathrm{SL(2,\bC)}$ Casimirs.
It leads to the condition
\be
\left(\rho - \gamma n\right)\left(\rho + \frac{n}{\gamma}\right) = 0\,,
\label{conditionfromcovariantconstraint}
\ee
on irreps of $\mathrm{SL(2,\bC)}$, and hence to $\rho = \gamma n$ or $\rho = -n/\gamma$.
Thus, we permit only irreps $\clH_{(\gamma n,n)}$ or $\clH_{(-n/\gamma,n)}$.

The constraints \eq{simplicityconstraintsaftergaugefixing}, on the other hand, are second--class,
and their quantization is more involved. They lead to the condition $4\gamma C_3 = C_2$ on irreps,
where $C_3$ is the Casimir of the little group defined by $U$. This constraint can be derived in three
different ways, following the three methods in the toy example above.

\subsection{Weak imposition of constraints}

Let us first consider the case $U = (1,0,0,0)$. The choice of gauge reduces the symmetry group from $\SLC$ to
SU(2), so we use the SU(2) decomposition \eq{decompositionSU(2)} of the kinematic Hilbert space $\clH_{(\rho,n)}$.
Suppose the physical Hilbert space is given by a subspace
\be
\Hphys = \bigoplus_{j\in \clJ} \clD_j\,,
\ee
where $\clJ$ is a subset of the total set of spins $\{ j | j \ge n/2\}$. We then require that
\be
\b\varphi | \v{C} | \psi\ket = 0\qquad\mbox{$\forall$ $|\varphi\ket, |\psi\ket\in \Hphys$.}
\label{weakimposition}
\ee
Unless $\Hphys$ is trivial, this implies, in particular, that for some $j \ge n/2$,
\be
\b\varphi | \v{C} | \psi\ket = 0\qquad\mbox{$\forall$ $|\varphi\ket, |\psi\ket\in \clD_j$.}
\ee
Therefore, one has
\be
\left\b j\,m' \left| J^3 + \frac{1}{\gamma} K^3 \right| j\,m\right\ket
= \left\b j\,m' \left| J^+ + \frac{1}{\gamma} K^+ \right| j\,m\right\ket
= \left\b j\,m' \left| J^- + \frac{1}{\gamma} K^- \right| j\,m\right\ket
= 0
\ee
for all admissible $m$, $m'$, with ladder operators given by
\be
J^\pm = J^1 \pm \irm J^2\qquad\mbox{and}\qquad K^\pm = K^1 \pm \irm K^2\,.
\ee
To analyze this, we use the action of $K^3$ on the canonical basis \cite{Carmeli}
\be
K^3 |j\, m\ket = (\ldots) |j+1\, m\ket - m A_j\, |j\, m\ket + (\ldots) |j-1\, m\ket\,,\qquad A_j = \frac{\rho\, n}{4 j(j+1)}\,,
\ee
and the commutation relation $K^\pm = \pm [K^3,J^\pm]$. It is then easy to see that all three equations imply $A_j = \gamma$,
which is equivalent to the aforementioned condition $4\gamma C_3 = C_2$.
In conjunction with the constraint $B\cdot\star B = 0$, this gives $4 j(j+1) = n^2$ if $\rho = \gamma n$ and $4 j(j+1) = -\rho^2$
if $n = -\gamma\rho$. The first case can be solved approximatively by $j = n/2$, while the second possibility gives a contradiction.
Therefore, one obtains that $\Hphys$ is only non--trivial when $\rho = \gamma n$ and then $\Hphys = \clD_{n/2}$.

Next we come to the case $U = (0,0,0,1)$. We impose again eq.\ \eq{weakimposition}, but this time with
the constraint $\v{C} = \v{F} + \frac{1}{\gamma}\v{G}$ and with regard to the SU(1,1) decomposition \eq{decompositionSU(1,1)}.
Suppose first that the constraint holds for some irrep $\clD^\pm_j$ of the discrete series. That is,
\be
\left\b j\,m' \left| F^0 + \frac{1}{\gamma} G^0 \right| j\,m\right\ket
= \left\b j\,m' \left| F^+ + \frac{1}{\gamma} G^+ \right| j\,m\right\ket
= \left\b j\,m' \left| F^- + \frac{1}{\gamma} G^- \right| j\,m\right\ket = 0\,.
\label{weakimpositionSU(1,1)}
\ee
Here,
\be
F^\pm = F^2 \mp \irm F^1\,,\qquad G^\pm = G^2 \mp \irm G^1\,,
\ee
and $F^\pm$  are the ladder operators of SU(1,1) \cite{LindbladNagel}\footnote{Note that Lindblad and Nagel use different symbols for the generators,
and in the discrete series their sign convention for $j$ is opposite to ours \cite{LindbladNagel}.}:
\be
F^\pm |j\, m\ket = \sqrt{(m\mp j \pm 1)(m\pm j)} |j\, m\pm 1\ket\,.
\ee
According to Mukunda\footnote{See eq.\ (3.19) in \cite{Mukunda} and also \cite{CHrep}.} the action of $K^3$ on SU(1,1) states is given by
\be
K^3 |j\, m\ket = (\ldots) |j+1\, m\ket - m \tilde{A}_j\, |j m\ket + (\ldots) |j-1\, m\ket\,,\qquad \tilde{A}_j = \frac{\rho\, n}{4 j(j-1)}\,.
\ee
Knowing that $G^\pm = \pm [G^0,F^\pm]$
we find, similarly as before, that the three equations \eq{weakimpositionSU(1,1)} imply $\tilde{A}_j = \gamma$ and hence $4\gamma C_3 = C_2$.
The solution is again $\rho = \gamma n$ and $j = n/2$. Since $j > 1/2$ in the decomposition \eq{decompositionSU(1,1)}, we also need that $n \ge 2$.

Assume finally that eq.\ \eq{weakimpositionSU(1,1)} holds for some irrep $\clC^\epsilon_s$ of the continuous series.
Then, we have the same equations except that $\tilde{A}_j$ is replaced by
\be
A_j = \frac{\rho\, n}{4 j(j+1)} = -\frac{\rho\, n}{4 (s^2 + 1/4)}\,.
\ee
A solution exists only when $\rho = -n/\gamma < -1$ and then
\be
s^2 + 1/4 = \frac{\rho^2}{4} = \frac{n^2}{4\gamma^2}\,.
\ee
The overall result for $U = (0,0,0,1)$ is that $\Hphys$ is only non--trivial when $\rho = \gamma n$, $n \ge 2$, or $\rho = -n/\gamma < -1$, and
in these cases
\be
\Hphys = \clD^+_{n/2} \oplus \clD^-_{n/2}\qquad\mbox{and}\qquad \Hphys = \clC^\epsilon_{\frac{1}{2}\sqrt{n^2/\gamma^2-1}} \oplus \clC^\epsilon_{\frac{1}{2}\sqrt{n^2/\gamma^2-1}}
\ee
respectively. In the continuous series, a subtlety arises from the fact that the states $|\Psi^{(\alpha)}_{s\, m}\ket$ are not normalizable.
If one wants to avoid singular inner products, a smearing w.r.t.\ $s$ is required.

\subsection{Restriction of coherent state basis}

\renewcommand{\arraystretch}{3}
The derivation from coherent states is essentially the one given in ref.\ \cite{CHtimelike},
but it is put more clearly in context with other methods by using the notion of a projector from
the kinematic to the physical Hilbert space.

When $U = (1,0,0,0)$, we resolve the identity on $\clH_{(\rho,n)}$ in terms of SU(2) coherent states:
\be
\mathbbm{1}_{(\rho,n)} = \sum\limits_{j = n/2}^\infty (2j + 1) \int\limits_{S^2} \d^2 N\; \left|\Psi_{j\v{N}}\right\ket \left\b\Psi_{j\v{N}}\right|\,.
\ee
The physical Hilbert space $\Hphys$ results from restricting the coherent state basis to states whose expectation values are
simple\footnote{The same method applies also to the first--class gauge constraint.
The projector on the gauge--invariant Hilbert space is equal to an integral over coherent states whose classical
labels (or expectation values) satisfy the closure constraint \cite{CFquantumgeometry}.}, i.e.\
to states for which
\be
\left\b\Psi_{j\v{N}}\right| \v{C} \left|\Psi_{j\v{N}}\right\ket = 0\,.
\ee
As shown in \cite{CHtimelike}, this implies $\gamma = A_j$.
With regard to the condition \eq{conditionfromcovariantconstraint}, we have two options. One can obtain \eq{conditionfromcovariantconstraint} either from the requirement that the
coherent state should have minimal uncertainty in $\v{K}$ (as done in \cite{CHtimelike}), or alternatively one can impose $B\cdot\star B = 0$ as
a separate constraint. The variance in $\v{K}$ equals
\be
\left(\Delta K\right)^2 = \b \v{K}^2 \ket - \b \v{K} \ket^2 = \b \v{J}^2\ket - \frac{1}{2}C_1 - \b \v{K} \ket^2\,.
\label{varianceK}
\ee
From $\gamma = A_j$ it follows that
\be
\b \v{J}^2\ket = \frac{1}{\gamma^2}\b \v{K}\ket^2 + O(|\v{J}|)\qquad\mbox{and}\qquad \b \v{J}^2 \ket = -\frac{1}{\gamma} \b \v{J}\cdot\v{K}\ket\,,
\ee
where $\b\; \ket$ indicates expectation values w.r.t.\ coherent states and $|\v{J}| \equiv \sqrt{|\b\v{J}\ket|}$.
By inserting the last two equations in \eq{varianceK} we arrive at
\bea
\left(\Delta K\right)^2 &=& -\frac{1}{\gamma}(1 - \gamma^2) \v{J}\cdot\v{K} - \frac{1}{2} C_1 + O(|\v{J}|) \\
&=& -\frac{\gamma}{4}\left[\left(1 - \frac{1}{\gamma^2}\right)C_2 + \frac{2}{\gamma} C_1\right] + O(|\v{J}|) \\
&=& -\frac{\gamma}{4} B\cdot\star B + O(|\v{J}|)\,,
\eea
which leads us back to the diagonal constraint.

Either way the solution is $\rho = \gamma n$ and $j = n/2$, so for $\rho = \gamma n$ the projector on the physical Hilbert space
becomes
\be
\Pphys = (n+1) \int\limits_{S^2} \d^2 N\; \left|\Psi_{n/2\v{N}}\right\ket \left\b\Psi_{n/2\v{N}}\right|\,.
\ee
For $U = (0,0,0,1)$, the little group is SU(1,1) and the relevant completeness relation is given by eq.\ \eq{totalcompletenessrelationSU(1,1)}.
Again, $\Hphys$ is defined by restricting to states whose expectation values are simple, i.e.
\be
\left\b\Psi^\pm_{j\, g}\right| \v{C} \left|\Psi^\pm_{j\, g}\right\ket = 0
\qquad\mbox{and}\qquad
\lim\limits_{\delta\to 0} \left\b\Psi^{(\alpha)}_{s\, g\, \delta}\right| \v{C} \left|\Psi^{(\alpha)}_{s\, g\, \delta}\right\ket = 0\,.
\ee
This implies $\gamma = \tilde{A}_j$ and $\gamma = A_j$ respectively, and the projector on the physical Hilbert space takes the form
\[
\Pphys^\delta
=
\left\{
\begin{array}{l@{\quad\mbox{if}\quad}l}
\ds (n-1) \int\limits_{\mathrm{SU(1,1)}}\!\!\!\! \d g\; \left(\left|\Psi^+_{n/2\, g}\right\ket \left\b\Psi^+_{n/2\, g}\right| + \left|\Psi^-_{n/2\, g}\right\ket \left\b\Psi^-_{n/2\, g}\right|\right)\,,
& \ds \rho = \gamma n\,, \\
\ds \int\limits_{\mathrm{SU(1,1)}}\!\!\!\! \d g\;
\sum_{\alpha = 1,2}\;\left|\Psi^{(\alpha)}_{\frac{1}{2}\sqrt{n^2/\gamma^2-1}\, g\, \delta}\right\ket \left\b\Psi^{(\alpha)}_{\frac{1}{2}\sqrt{n^2/\gamma^2-1}\, g\, \delta}\right|\,,
& \ds \rho = - n/\gamma < -1\,.
\end{array}
\right.
\]

\subsection{Master constraint}

The master constraint is the original method by which the simplicity constraints
of the EPRL model were derived \cite{EPRL}. However, the same technique can be also
applied to determine simplicity constraints for a spacelike normal $U$.

The master constraint is given by the sum of the squares of the components of the simplicity constraint.
For $U = (1,0,0,0)$, this yields
\be
M = \left(1 + \frac{1}{\gamma^2}\right) \v{J}^2 - \frac{1}{2\gamma^2} C_1 - \frac{1}{2\gamma} C_2 = 0\,.
\label{masterconstraintSU(2)}
\ee
The diagonal constraint $B\cdot \star B = 0$, on the other hand, is equivalent to
\be
\left(1 - \frac{1}{\gamma^2}\right) C_2 + \frac{2}{\gamma} C_1 = 0\,.
\label{covariantconstraintv2}
\ee
By combining the diagonal and master constraint, one arrives at the desired second condition
\be
4\gamma \v{J}^2 = 4\gamma C_3 = C_2\,.
\label{J2C2}
\ee
Similarly, simplicity for $U = (0,0,0,1)$ results in the master constraint
\be
M = \left(1 + \frac{1}{\gamma^2}\right) \v{F}^2 - \frac{1}{2\gamma^2} C_1 - \frac{1}{2\gamma} C_2 = 0\,.
\ee
After combination with \eq{covariantconstraintv2}, one has $4\gamma \v{F}^2 = 4\gamma C_3 = C_2$.
Therefore, one obtains the same overall solution as with the previous two techniques.

A point of concern could be the fact that in the SU(1,1) case
the master constraint is not positive definite.
For SU(2) the master constraint is always positive and hence
the vanishing of $M$ implies that each individual constraint
vanishes classically.
This reasoning does not apply to the SU(1,1) case,
so one might worry that additional conditions are needed
when using the master constraint.
In the previous two subsections, however, exactly the same constraints
followed from considerations that involved each individual simplicity
constraint. This suggests that the master and diagonal constraint
are sufficient and that the indefinite sign does not cause any problems.

\section{Spin foam sum as an integral over coherent states}
\label{coherentstateintegral}

\renewcommand{\arraystretch}{2}
With the completeness relations of \sec{completenessrelations} at hand, we are ready to describe the spin foam model \cite{CHtimelike}
in terms of integrals over coherent states.
For this purpose, it is convenient to use a uniform notation for each of the different cases occurring in the quantization.
We have to distinguish between the two different possibilities $U = (1,0,0,0)$ and $U = (0,0,0,1)$ for the normals of tetrahedra,
and furthermore between spacelike and timelike triangles within such tetrahedra. The choice between spacelike and timelike
triangles is indicated by the variable $\zeta = \pm 1$.

In the quantum theory, triangles are represented by states in unitary irreps $\clH_{(\rho,n)}$ of $\SLC$. The simplicity constraints
require that $\rho = \gamma n$, $n\ge 2$, for spacelike triangles and $\rho = -n/\gamma$, $n > \gamma$, for timelike triangles.
For $U = (1,0,0,0)$, the tetrahedral space is Euclidean and
triangles are spacelike. These triangles correspond to coherent states in the SU(2) decomposition \eq{decompositionSU(2)}.
When $U = (0,0,0,1)$, the tetrahedron resides in a 3d Minkowski space and a triangle can be spacelike or timelike;
the associated coherent state belongs to the discrete or continuous series of the SU(1,1) decomposition \eq{decompositionSU(1,1)}
respectively.

The coherent states are parametrized by two labels. The first label is a spin $j$; it determines the area eigenvalue of the triangle.
The second label is an element of SU(2) or SU(1,1), or alternatively
a unit 3--vector $\v{N}$ in the 2--sphere or the 2d hyperboloids.
Geometrically, the vector $\v{N}$ has the meaning of a normal vector to a triangle.
In a Minkowskian tetrahedron, $\v{N}$ is timelike when the triangle is spacelike, and
spacelike when the triangle is timelike. The spin $j$ is subject to the simplicity constraints
\be
j = \left\{
\begin{array}{l@{\mbox{\qquad if\qquad}}ll}
n/2\,, & \zeta = 1\,, & U = (1,0,0,0)\,, \\
n/2\,, & \zeta = 1\,, & U = (0,0,0,1)\,, \\
-\frac{1}{2} + \frac{\irm}{2}\sqrt{n^2/\gamma^2 - 1}\,, & \zeta = -1\,, & U = (0,0,0,1)\,. \\
\end{array}
\right.
\label{simplicityj}
\ee
The corresponding area spectra (see \cite{CHtimelike}) are given by
\be
A = \left\{
\begin{array}{l@{\mbox{\qquad if\qquad}}ll}
\gamma \sqrt{j(j+1)}\,, & \zeta = 1\,, & U = (1,0,0,0)\,, \\
\gamma \sqrt{j(j-1)}\,, & \zeta = 1\,, & U = (0,0,0,1)\,, \\
\gamma\sqrt{s^2 + 1/4} = n/2\,, & \zeta = -1\,, & U = (0,0,0,1)\,. \\
\end{array}
\right.
\ee
Below the different choices of the little group are subsumed in the formula
\be
H(\zeta,U) \equiv \left\{
\begin{array}{l@{\mbox{\qquad if\qquad}}ll}
\mathrm{SU(2)}\,, & \zeta = 1\,, & U = (1,0,0,0)\,, \\
\mathrm{SU(1,1)}\,, & \zeta = \pm 1\,, & U = (0,0,0,1)\,, \\
\varnothing\,, & \zeta = -1\,, & U = (1,0,0,0)\,.
\end{array}
\right.
\label{uniformH}
\ee
Moreover, coherent states are uniformly written as $|\Psi^{(\alpha)}_{j\, h\, \delta}\ket$, where $h\in H(\zeta,U)$. In the SU(2) case, $\alpha$ can assume
only one value and the state is equal to an SU(2) coherent state, i.e.\ $|\Psi^{(\alpha)}_{j\, h\, \delta}\ket \equiv |\Psi_{j\, h}\ket$ and $h\in \mathrm{SU(2)}$.
In the second case of eq.\ \eq{uniformH}, the state is a coherent state of the discrete series,
$|\Psi^{(\alpha)}_{j\, h\, \delta}\ket \equiv |\Psi^\alpha_{j\, h}\ket$, and $\alpha$ can be $+$ or $-$. In the third case, it
is a coherent state of the continuous series and $\alpha = 1, 2$.
The right--hand side of the completeness relations \eq{completenessSU(2)}, \eq{completenessdiscrete} and \eq{completenesscontinuousg} may now be cast in the form
\be
d_n(\zeta,U) \sum_\alpha \int\limits_{H(\zeta,U)} \!\!\!\d h\; \left|\Psi^{(\alpha)}_{j\, h\, \delta}\right\ket \left\b \Psi^{(\alpha)}_{j\, h\, \delta}\right|\,,
\label{unifiedcompleteness}
\ee
where
\be
d_n(\zeta,U) = \left\{
\begin{array}{l@{\mbox{\qquad if\qquad}}ll}
n + 1\,, & \zeta = 1\,, & U = (1,0,0,0)\,, \\
\theta(n-2) (n - 1)\,, & \zeta = 1\,, & U = (0,0,0,1)\,, \\
1 - \theta(\gamma - n)\,, & \zeta = -1\,, & U = (0,0,0,1)\,, \\
0\,, & \zeta = -1\,, & U = (1,0,0,0)\,.
\end{array}
\right.
\ee
The Heaviside functions are inserted to ensure that the expression is zero unless $n \ge 2$ in the case of the discrete class states
and $n > \gamma$ in the case of the continuous series.

The partition function of spin foams is defined on a 4--dimensional simplicial complex $\Delta$
and its dual complex $\Delta^*$. We denote edges, triangles, tetrahedra and 4--simplices of $\Delta$ by $l$, $t$, $\tau$ and $\sigma$ respectively.
For dual vertices, edges and faces we use $v$, $e$ and $f$. Note that dual edges $e$ and faces $f$ stand in one--to--one correspondence
with tetrahedra $\tau$ and triangles $t$ of the original complex.

Configurations are specified by the following data. To each face $f$ one assigns a positive integer $n_f$ which determines $\rho_f$ and the $\SLC$ irrep $(\rho_f,n_f)$
via the simplicity constraints. For each face and edge, there are labels $\zeta_f$ and $U_e$ that specify the signature of the corresponding
triangle and tetrahedron.
To each edge $e$ and adjacent face $f$, we attribute a coherent state $|\Psi^{(\alpha_{ef})}_{j_{ef} h_{ef}\delta}\ket$.
Equivalently, one can use states that are parametrized by unit 3--vectors $\v{N}$ in the 2--sphere $S^2$ and the hyperboloids $\bH_\pm$ and $\bH_{\mathrm{sp}}$.
(Below we adopt the labeling with group elements, since relation \eq{completenesscontinuousg} requires less notation than \eq{completenesscontinuousN}.)
Furthermore, there are connection variables $g_{ev}\in \SLC$ assigned to half--edges of $\Delta^*$, going from the vertex $v$ to the center of the edge $e$.

The completeness relations are associated to edges $e$. When the amplitude is organized in terms of vertices $v$, the
completeness relations are split in halves in the sense that the ket state goes to one end of the edge and the bra state goes to the other end.
The result are vertex amplitudes of the form
\be
A_v((\rho_f,n_f);h_{ef},\alpha_{ef},\delta) = \int\limits_{\mathrm{SL(2,\bC)}} \prod_e \d g_{ev}\; \prod_f
\left\b \Psi^{(\alpha_{ef})}_{j_{ef} h_{ef}\delta} \right| D^{(\rho_f,n_f)}(g_{ev} g_{ve'}) \left| \Psi^{(\alpha_{e'\!f})}_{j_{e'\!f} h_{e'\!f}\delta}\right\ket\,.
\ee
Each face (adjacent to the vertex) contributes a factor that results from the inner product of two coherent states belonging to the edges
$e$ and $e'$ adjacent to the face $f$. The vertex amplitude is obtained by integrating the product of these factors over the connection variables
$g_{ev}$ on half--edges.

The partition function is given by a multiple integral over vertex amplitudes:
\be
Z = \sum_{n_f} \sum_{\zeta_f=\pm 1} \sum_{U_e} \sum_{\alpha_{ef}} \int\limits_{H(U_e,\zeta_f)}\!\!\!\!\!\! \d h_{ef}\,d_{n_f}(U_e,\zeta_f)
\;\prod_f (1 + \gamma^{2\zeta_f}) n_f^2\;
\lim_{\delta\to 0}\,\prod_v A_v\!\left((\zeta_f\gamma^{\zeta_f} n_f,n_f);h_{ef},\alpha_{ef},\delta\right)
\ee
First of all, there is a sum over the various discrete labels $n_f$, $\zeta_f$, $U_e$ and $\alpha_{ef}$.
For each edge $e$ and adjacent face $f$, one has an integral over the subgroup $H(U_e,\zeta_f)$ whose elements $h_{ef}$ parametrize the coherent states.
Each face comes with a measure factor that descends from the original measure in the unconstrained BF theory.
Finally, there is a limit $\delta\to 0$ on the product of all vertex amplitudes.
Physically, the parameter $\delta$ has the meaning of an uncertainty in area for timelike triangles,
and it is introduced to avoid singular inner products between states of the continuous series.
Once the inner products are computed, this parameter is sent to zero.

It should be remarked that the simplicity constraint permit only certain combinations of states
around faces. If a triangle is timelike, it carries only continuous series
states in the adjacent tetrahedra. In contrast, spacelike triangles admit
both SU(2) states and SU(1,1) states of the discrete series.

\section{Discussion}
\label{discussion}

Let us summarize our results.
We dealt with the spin foam theory of ref.\ \cite{CHtimelike}---an
extension of the EPRL model to triangulations that contain both Euclidean and Lorentzian tetrahedra
(and hence both spacelike and timelike triangles).
We expressed its partition function as a multiple integral over vertex amplitudes
that have coherent states as boundary data.
Each coherent state is interpreted as a quantum state of a triangle and
it is characterized by a spin and a unit 3--vector. This vector is the normal of the triangle
and lies either in the 2--sphere or the 2d hyperboloids, depending on the signature of the tetrahedral space.
In order to write the partition function in terms of vertex amplitudes,
we required completeness relations for coherent states in each of the irreps of SU(2) and SU(1,1).
In the case of the continuous series, a new type of coherent state was employed
and a corresponding completeness relation was established.

We demonstrated furthermore that the physical Hilbert space of the simplicity constraints
can be derived by three different methods:
by the weak imposition of constraints, by the master constraint (as advocated in the EPR and EPRL papers \cite{EPR,EPRflipped,EPRL}),
and by the restriction of coherent state bases (inspired by the FK model \cite{FK,Livine_Consistently}).
The agreement of these three techniques supports the idea that the result is correct.
However, quantization rules are just rules of thumb and the true test comes
when the physical behavior of the system is investigated.

One way to check this is to determine the large spin asymptotics of vertex amplitudes.
This was already done for the Riemannian and Lorentzian EPRL model \cite{CFpathrep,CFsemiclassical,BarrettasymptoticsEuclidean,BarrettasymptoticsLorentzian,BarrettasymptoticsEverything},
and it may be possible to extend the same analysis to the present form of the theory.
This may require some technical effort, as we have to deal with the
SU(1,1) reduction of $\SLC$ representations, which is less explored than the
canonical SU(2) decomposition.
There will appear, in particular, inner products between SU(2) states and
SU(1,1) states of the discrete series.
It would be interesting to understand the relation between these two kind of states,
since they represent the same type of triangle in tetrahedra of different signature.

The most interesting aspect of the model is the fact that it admits
timelike boundaries. One may now consider finite regions with spacelike and
timelike boundaries, as envisioned in papers by Rovelli et al.\ \cite{RovellietalMinkowskivacuum}
and Oeckl \cite{OecklCat,Oecklgeneralboundary}.
Connected to this, there is the question of whether one can have some form
of canonical loop quantum gravity for states on timelike boundaries (see
\cite{AlexandrovKadar} for earlier work on this).
A well--known issue in this regard is the absence of the trivial representation
in non--compact gauge groups, which makes it difficult to relate Hilbert spaces of
different graphs. As long as one uses a single graph, however, this may not be a major
obstacle.

When defining the spin foam sum, we were mainly guided by the covariant picture and
by the idea that we should sum over all geometries. For this reason, we summed
over the two choices of the tetrahedral normal. From the canonical perspective, however,
one would also require that the sum over intermediate states on a triangulation's boundary
is a projector. For a given choice of the normal, this is certainly the case.
One has to check if this is still true when we sum over normals, or
if one needs a restriction on states to ensure the projection property.
More on this will be reported elsewhere.

\begin{acknowledgments}
I thank Sergei Alexandrov, Eugenio Bianchi, Laurent Freidel, Jeff Hnybida,  Jonathan McDonald and Roberto Pereira for discussions.
Research at Perimeter Institute is supported by the Government of Canada through Industry Canada and by the Province of Ontario
through the Ministry of Research \& Innovation.
\end{acknowledgments}

\begin{appendix}

\section{Proof of completeness relations}
\label{proof}

Below we prove the completeness relations \eq{completenesscontinuousg} and \eq{completenesscontinuousN} for coherent states
in the continuous series. A general state in the continuous series subspace of $\clH_{(\rho,n)}$ can be written as
\be
|u\ket = \sum_{\alpha = 1,2} \sum_{\pm m = \epsilon} \int\limits_0^\infty \d s\; \sqrt{\mu_\epsilon(s)}\, u^{(\alpha)}_m(s)\left|\Psi^{(\alpha)}_{s'\, m}\right\ket\,.
\ee
Here, $u^{(\alpha)}_m(s)$ plays the role of a wavefunction. The inner product of two such states $|u\ket$ and $|v\ket$ yields
\be
\b u | v\ket = \sum_{\alpha = 1,2} \sum_{\pm m = \epsilon} \int\limits_0^\infty \d s\; u^{(\alpha) *}_m(s) v^{(\alpha)}_m(s)\,.
\ee
When the projector \eq{projector_s} is sandwiched between $|u\ket$ and $|v\ket$, we obtain
\be
\b u | P^\epsilon_s(\delta) | v\ket
=
\sum_{\alpha = 1,2} \sum_{\pm m = \epsilon} \int\limits_0^\infty \d s'\; u^{(\alpha) *}_m(s') f_\delta(s'-s) v^{(\alpha)}_m(s')\,.
\ee
We would like to show that the coherent state integral on the right--hand side of \eq{completenesscontinuousg}
produces the same after sandwiching between $|u\ket$ and $|v\ket$.
When all smearing integrals are written explicitly, this contraction
gives
\bea
\lefteqn{\sum_{\alpha = 1,2}\; \int\limits_{\mathrm{SU(1,1)}}\!\!\!\! \d g\;
\left\b u \left|\Psi^{(\alpha)}_{s\, g\, \delta}\right\ket\right. \left.\left\b\Psi^{(\alpha)}_{s\, g\, \delta}\right|v\right\ket} \nonumber \\
&=&
\sum_{\alpha_1 = 1,2} \sum_{\pm m_1 = \epsilon} \int\limits_0^\infty \d s_1\; \sqrt{\mu_\epsilon(s_1)}\, u^{(\alpha) *}_m(s_1)
\sum_{\alpha_2 = 1,2} \sum_{\pm m_2 = \epsilon} \int\limits_0^\infty \d s_2\; \sqrt{\mu_\epsilon(s_2)}\, v^{(\alpha)}_m(s_2) \nonumber \\
&&
\times\,
\int\limits_0^\infty \d s'_1\; \mu_\epsilon(s'_1)\, f_\delta(s'_1-s)
\int\limits_{-\infty}^\infty \d\lambda_1\; \frac{1}{\sqrt{\delta}} f_\delta(\lambda_1 - s)
\int\limits_0^\infty \d s'_2\; \mu_\epsilon(s'_2)\, f_\delta(s'_2-s)
\int\limits_{-\infty}^\infty \d\lambda_2\; \frac{1}{\sqrt{\delta}} f_\delta(\lambda_2 - s) \nonumber \\
&&
\times\,
\sum_{\alpha = 1,2}\, \int\limits_{\mathrm{SU(1,1)}}\!\!\!\! \d g\;
\left\b\Psi^{(\alpha_1)}_{s_1\,m_1}\right|D^{(\rho,n)}(g)\left|\Psi^{(\alpha)}_{s'_1\,\lambda_1}\right\ket
\left\b\Psi^{(\alpha)}_{s'_2\,\lambda_2}\right|D^{(\rho,n)}(g^{-1})\left|\Psi^{(\alpha_2)}_{s_2\,m_2}\right\ket\,.
\eea
Because of the isomorphism \eq{isomorphismcontinuousseries} with states in $\clC^\epsilon_s$ this reduces to
\bea
\lefteqn{\sum_{\alpha = 1,2}\; \int\limits_{\mathrm{SU(1,1)}}\!\!\!\! \d g\;
\left\b u \left|\Psi^{(\alpha)}_{s\, g\, \delta}\right\ket\right. \left.\left\b\Psi^{(\alpha)}_{s\, g\, \delta}\right|v\right\ket} \nonumber \\
&=&
\sum_{\alpha = 1,2}
\sum_{\pm m_1 = \epsilon} \int\limits_0^\infty \d s_1\; \sqrt{\mu_\epsilon(s_1)}\, u^{(\alpha) *}_m(s_1) f_\delta(s_1-s)
\int\limits_{-\infty}^\infty \d\lambda_1\; \frac{1}{\sqrt{\delta}} f_\delta(\lambda_1 - s) \nonumber \\
&&
\times\,
\sum_{\pm m_2 = \epsilon} \int\limits_0^\infty \d s_2\; \sqrt{\mu_\epsilon(s_2)}\, v^{(\alpha)}_m(s_2) f_\delta(s_2-s)
\int\limits_{-\infty}^\infty \d\lambda_2\; \frac{1}{\sqrt{\delta}} f_\delta(\lambda_2 - s) \nonumber \\
&&
\times\,
\int\limits_{\mathrm{SU(1,1)}} \d g\;
\b j_1\,m_1|D^{j_1}(g)|j_1\,\lambda_1\ket
\b j_2\,\lambda_2|D^{j_2}(g^{-1})|j_2\,m_2\ket\,,
\eea
where $j_1 = -\frac{1}{2} + \irm s_1$ and $j_2 = -\frac{1}{2} + \irm s_2$.
By applying the orthogonality relation \eq{orthogonalitymatrixcontinuous} between matrix elements of SU(1,1), we arrive finally at
\bea
\lefteqn{\sum_{\alpha = 1,2}\; \int\limits_{\mathrm{SU(1,1)}}\!\!\!\! \d g\;
\left\b u \left|\Psi^{(\alpha)}_{s\, g\, \delta}\right\ket\right. \left.\left\b\Psi^{(\alpha)}_{s\, g\, \delta}\right|v\right\ket} \nonumber \\
&=&
\sum_{\alpha = 1,2} \sum_{\pm m = \epsilon} \int\limits_0^\infty \d s'\; u^{(\alpha) *}_m(s') f^2_\delta(s'-s)  v^{(\alpha)}_m(s')
\int\limits_{-\infty}^\infty \d\lambda\; \frac{1}{\delta} f^2_\delta(\lambda - s) \nonumber \\
&=&
\sum_{\alpha = 1,2} \sum_{\pm m = \epsilon} \int\limits_0^\infty \d s'\; u^{(\alpha) *}_m(s') f_\delta(s'-s) v^{(\alpha)}_m(s')\,.
\eea
This proves the completeness relation \eq{completenesscontinuousg}.

For the transition from SU(1,1) to the hyperboloid $\clH_{\mathrm{sp}}$,
the smearing in $s$ does not play any role, so we indicate the associated integrals
only by $\int\d s \ldots$. Starting from
\be
P^\epsilon_s(\delta) = \sum_{\alpha = 1,2}\; \int\limits_{\mathrm{SU(1,1)}}\!\!\!\! \d g\; \left|\Psi^{(\alpha)}_{s\, g\, \delta}\right\ket \left\b\Psi^{(\alpha)}_{s\, g\, \delta}\right|
\ee
we use the parametrization \eq{parametrizationcontinuous} and get
\bea
\lefteqn{P^\epsilon_s(\delta) =
\sum_{\alpha = 1,2}
\int\limits_0^\infty \d s_1 \ldots
\int\limits_{-\infty}^\infty \d\lambda_1\; \frac{1}{\sqrt{\delta}} f_\delta(\lambda_1 - s)
\int\limits_0^\infty \d s_2 \ldots
\int\limits_{-\infty}^\infty \d\lambda_2\; \frac{1}{\sqrt{\delta}} f_\delta(\lambda_2 - s)} \nonumber \\
&&
\hspace{-1cm}\times\,
\frac{1}{(4\pi)^2} \int\limits_{-2\pi}^{2\pi} \d\varphi \int\limits_{-\infty}^\infty \d t \int\limits_{-\infty}^\infty \d u\,\cosh t\;
\e^{\irm u(\lambda_1 - \lambda_2)}\,
D^{j_1}(g(\varphi,t,0)) \left|\Psi^{(\alpha)}_{s_1\,\lambda_1\, +}\right\ket \left\b \Psi^{(\alpha)}_{s_2\,\lambda_2\, +}\right| D^{j_2}(g^{-1}(\varphi,t,0)) \nonumber \\
&=&
\sum_{\alpha = 1,2}
\frac{4\pi}{(4\pi)^2} \int\limits_{-\pi}^{\pi}\d\varphi \int\limits_{-\infty}^\infty \d t\,\cosh t
\int\limits_0^\infty \d s_1 \ldots \int\limits_0^\infty \d s_2 \ldots
\int\limits_{-\infty}^\infty \d\lambda\; \frac{1}{\delta} f_\delta(\lambda - s) \nonumber \\
&&
\times\,
D^{j_1}(g(\varphi,t,0))\, \left|\Psi^{(\alpha)}_{s_1\,\lambda\, +}\right\ket \left\b \Psi^{(\alpha)}_{s_2\,\lambda\, +}\right| D^{j_2}(g^{-1}(\varphi,t,0)) \\
&=&
\sum_{\alpha = 1,2}\; \int\limits_{\bH_{\mathrm{sp}}} \d^2 N
\int\limits_{-\infty}^\infty \d\lambda\; \frac{1}{\delta} f_\delta(\lambda - s)\,
\left|\Psi^{(\alpha)}_{j\v{N}\lambda\delta}\right\ket \left\b\Psi^{(\alpha)}_{j\v{N}\lambda\delta}\right|\,.
\eea
Since $\e^{2\pi \irm J^3}$ produces the same sign factor for bra and ket, we were able to restrict $\varphi$ to $(-\pi,\pi]$ and compensate by a factor of 2.
This corresponds to the division by $\bZ_2$ in the quotient $\mathrm{SU(1,1)} / (\mathrm{G_1}\otimes\bZ_2) \simeq \bH_{\mathrm{sp}}$.

\end{appendix}

\end{document}